\documentclass[9pt,twocolumn,twoside]{opticajnl}
\journal{opticajournal} 

\setboolean{shortarticle}{true}

\usepackage{lineno}

\usepackage{siunitx}

\title{Soliton generation through temporal reflection in media with a frequency-dependent nonlinearity}

\author[1,2,*]{Lucas N. Gutierrez}
\author[2,3]{Pablo I. Fierens}
\author[1,2]{Diego F. Grosz}
\author[1,2]{Santiago M. Hernandez}

\affil[1]{Grupo de Comunicaciones Ópticas, Depto. de Ingenier\'ia en Telecomunicaciones, Instituto Balseiro, Comisi\'on Nacional de Energ\'ia At\'omica, R\'io Negro 8400, Argentina}
\affil[2]{Consejo Nacional de Investigaciones Cient\'ificas y T\'ecnicas (CONICET), CABA 1425, Argentina}
\affil[3]{Instituto Tecnol\'ogico de Buenos Aires (ITBA), CABA 1437, Argentina}

\affil[*]{lucas.gutierrez@ib.edu.ar}

\begin{abstract}

We demonstrate that the temporal reflection of a weak dispersive pulse on a soliton in media with a frequency-dependent nonlinearity leads to the generation of new solitons, whose number can be selected by tuning parameters of the dispersive pulse. By carefully analyzing the different processes involved, we show that a virtuous interplay between Raman scattering and a zero-nonlinearity wavelength
is a key enabler for soliton generation to occur, limiting the initial soliton redshift and allowing for an efficient energy transfer between the dispersive pulse and its reflection.
Finally, we believe results presented in this work may contribute guidelines for the generation of short and intense pulses for various photonic applications.

\end{abstract}

\setboolean{displaycopyright}{false} 

\begin{document}

\maketitle

The interaction between dispersive pulses and fundamental solitons is a topic that has attracted considerable attention in the past decade. Such a nonlinear interaction has been shown to find applications in all-optical switching and control~\cite{Pickartz:16_All_Optical},  supercontinuum generation~\cite{Demircan:13_Compressible_SC, Demircan:14_Multiple_Scatterings}, and even applied to explain phenomena like rogue waves~\cite{Demircan:13_Rogue}.

Waveguides made of media with a frequency-dependent nonlinear response have also been of particular interest recently. These waveguides can be realized through, for instance, by doping the core of a photonic crystal fiber with silver nanoparticles~\cite{Bose:16_Implications_ZNW}. Most interestingly, such media may exhibit a zero-nonlinearity wavelength (ZNW) where the nonlinear response vanishes~\cite{Sparapani:2024:freqdependent}. The presence of a ZNW may lead to the enhancement or suppression of the Raman redshift of fundamental solitons~\cite{Arteaga:18_Soliton_Freq_Kerr} and has been shown to allow for an efficient all-optical control~\cite{Sparapani:24_All_Optical}. 

In this work, we study the interaction between a dispersive pulse and a fundamental soliton in a waveguide with a ZNW. We show how this interaction can lead to the generation of new intense short pulses, depending upon the initial parameters of the dispersive pulse as well as the actual position of the ZNW. Note that for this study, we cannot rely on the generalized nonlinear Schrödinger equation, commonly used to model nonlinear propagation~\cite{Agrawal:19_Nonlinear}, as it has been shown to predict unphysical results such as the failure to conserve the photon number in lossless waveguides with a frequency-dependent nonlinearity. To address this issue, a modified equation, known as the photon-conserving generalized nonlinear Schrödinger equation (pcGNLSE)~\cite{bonetti:20_pcgnlse}, was introduced. In the frequency domain the pcGNLSE reads
\begin{equation}
\begin{split}
    \partial_z \tilde A & =  i \beta(\Omega)\tilde A + i \frac{\Gamma(\Omega)}{2}\mathcal{F}\{C^* B^2\} + i \frac{\Gamma^*(\Omega)}{2}\mathcal{F}\{B^* C^2\}\\ &+ i f_R \Gamma^*(\Omega)\mathcal{F}\left\{B \int_0^\infty h_R(t') |B(t-t')|^2 - B |B|^2\right\},
\end{split}
\label{eq:pcgnlse}
\end{equation}
where $\Omega = \omega - \omega_0$ is the deviation from a central frequency $\omega_0$, $\tilde A(z,\Omega)$ is the Fourier transform of the complex envelope of the electrical field, $\beta(\Omega)$ is the dispersion profile, $h_R(t)$ the delayed Raman response, $f_R$ the fractional Raman contribution, and $\mathcal{F}$ the Fourier-transform operator. The auxiliary fields $B$ and $C$ are defined as $r(\Omega) \tilde A (z,\Omega)$ and $r^*(\Omega) \tilde A(z,\Omega)$, respectively, with $r(\Omega) = \sqrt[4]{\gamma(\Omega)\cdot (\omega_0 + \Omega)^{-1}}$ and $\gamma(\Omega)$ is the usual nonlinear parameter~\cite{Agrawal:19_Nonlinear}. The modified nonlinear profile is $\Gamma(\Omega)= \sqrt[4]{\gamma(\Omega) \cdot (\omega_0 + \Omega)^2}$. For the sake of simplicity, in this letter we assume a linear frequency-dependence of the frequency parameter, e.g., $\gamma(\Omega) = \gamma_0 + \gamma_1 \Omega$, and a cubic frequency dispersion profile $\beta(\Omega)=\beta_2\Omega^2/2+\beta_3\Omega^3/6$.

As it will be shown, the soliton spectral shift caused by its collision with the dispersive pulse plays a fundamental role in the processes leading to the generation of new solitons. In order to study such spectral shift, we launch a fundamental soliton at $\lambda_0 = \SI{1600}{\nano\meter}$, with peak power $P_0 = \SI{243}{\watt}$ and time width $T_0 = \SI{85}{\femto\second}$, into a \SI{300}{\meter}-long fiber with $\beta_2 = \SI{-4.4}{\pico\second^2\kilo\meter^{-1}}$, $\beta_3 = \SI{0.13}{\pico\second^2\kilo\meter^{-1}}$, $\gamma_0 = \SI{2.5}{\watt^{-1}\kilo\meter^{-1}}$, and ZNW $=\SI{1650}{\nano\meter}$. For the sake of clarity, at this point we ignore the effect of Raman scattering. Observe that the resulting zero-dispersion wavelength (ZDW) is \SI{1555.3}{\nano\meter}. A \textit{sech}-like unchirped dispersive pulse with peak power $P_\mathrm{d} = \SI{20}{\watt}$ and time width $T_\mathrm{d} = \SI{3}{\pico\second}$ was launched at $\lambda_\mathrm{d} = \SI{1500}{\nano\meter}$, initially delayed by $\Delta T = \SI{-20}{\pico\second}$. The chosen wavelength ensures a small group-velocity difference between the pulses, resulting in a high reflectivity for the dispersive pulse~\cite{Plansinis:15_Temporal_Analog}. 

\begin{figure}[t!]
\centering\includegraphics[width=\linewidth]{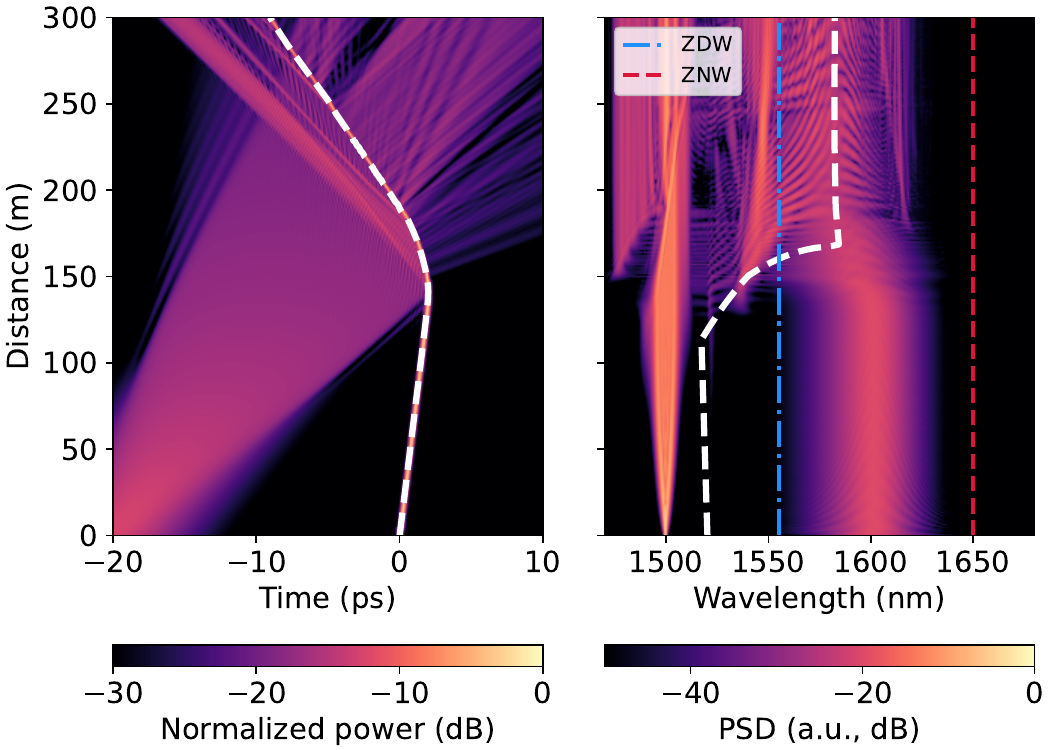}
\caption{Interaction between a dispersive pulse and a fundamental soliton in time (left) and frequency (right) domains. Dashed lines show the soliton trajectory fitted with a polynomial spline (left), and the predicted reflected wavelength (right).}
\label{fig:recoil}
\end{figure}

Figure~\ref{fig:recoil} shows the evolution in time (left) and frequency (right). The observed trajectory of the fundamental soliton can be explained by using the temporal reflection analogy~\cite{Plansinis:15_Temporal_Analog}. During the interaction, a portion of the dispersive pulse is reflected and shifts to a different frequency. Meanwhile, the soliton may be accelerated or decelerated and, if the dispersive pulse is sufficiently intense, the soliton can shift to longer or shorter wavelengths. The frequency of the reflected pulse can be estimated by the Doppler shift relation between incident and reflected pulses, by assuming the soliton  to be a moving barrier with a group velocity  $v = 1/\beta'(\Omega_s)$, where~\cite{Pickartz:16_Adiabatic} 
\begin{equation}
    \beta'(\Omega_s) = \frac{\beta(\Omega_\mathrm{Re}) - \beta(\Omega_\mathrm{In})  }{\Omega_\mathrm{Re} - \Omega_\mathrm{In}}.
\label{eq:reflection_estimation}
\end{equation}
Here, $\Omega_\mathrm{In}$ is the frequency of the incident dispersive pulse, $\Omega_\mathrm{Re}$ the frequency of the reflected pulse, and $\Omega_s$ is the frequency of the soliton. This relation allows for the estimation of the reflected frequency, even if the frequency of the soliton has shifted from its original value. 

The trajectory of the fundamental soliton was fitted with a polynomial spline in order to estimate its delay as a function of $z$ and, by taking the derivative, its group velocity. Using \eqref{eq:reflection_estimation} we can compute the reflected wavelength throughout propagation, which is shown as a dashed white line in the frequency domain. It is important to note that, during the interaction, there is no unique reflected wavelength~\cite{Demircan:13_Compressible_SC}; instead, reflected components are observed to span from $\lambda_\mathrm{d}$ to nearly $\lambda_0$. Moreover, as \eqref{eq:reflection_estimation} has no real roots for $z > \SI{170}{\meter}$ there is no reflection beyond that point. 

\begin{figure}[b!]
\centering\includegraphics[width=0.9\linewidth]{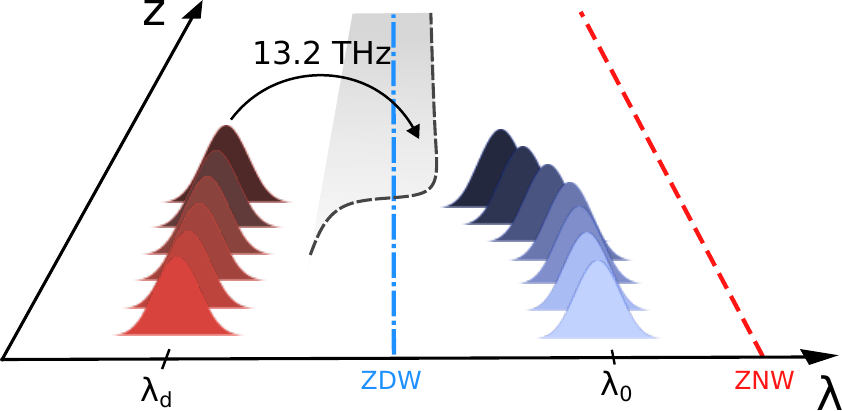}
\caption{Schematic diagram of pulse interaction in the frequency domain. The shaded region depicts the range where spectral content is generated through the temporal reflection. During the process, and in continuous fashion, the incident pulse transfers energy efficiently to the reflected pulse via Raman scattering.}
\label{fig:diagram}
\end{figure}

Prior to the collision, the estimated reflected wavelength lies within the normal dispersion regime where no solitons can be formed. As the pulses interact, the blueshift of the fundamental soliton causes the reflected wavelength to change continuously, crossing the ZDW into the solitonic regime. 
Figure~\ref{fig:recoil} shows that the reflected components that lie in the solitonic region do not exhibit enough power to evolve into solitons. 
One may consider increasing the power of the colliding dispersive pulse; however, this may lead to a rapid spectral shift of the soliton and thus losing the reflection condition, i.e., \eqref{eq:reflection_estimation} has no longer a real solution. Therefore, the power of the dispersive pulse does not lend itself as a simple control parameter enabling or inhibiting the emergence of new solitons. This way, we turn to another mechanism that allows an energy transfer from the incident dispersive pulse to reflected components such as the Raman gain. Although the selection of an adequate incident wavelength is required for an efficient energy transfer (as shown in Figure~\ref{fig:diagram}), the fact that there is a broad reflected band provides margin for tuning.

Following these considerations, we change the incident pulse wavelength to $\lambda_\mathrm{d} = \SI{1480}{\nano\meter}$. Briefly postponing the analysis of their influence, we also slightly modify the dispersive pulse power and width to $P_\mathrm{d} = \SI{30}{\watt}$ and $T_\mathrm{d} = \SI{5.75}{\pico\second}$. The delayed Raman response is modeled as~\cite{Agrawal:19_Nonlinear}
\begin{equation}
    h_{\mathrm{R}}(t) = \frac{\tau_1^2 + \tau_2^2}{\tau_1 \tau_2^2} \sin(t/\tau_1) \mathrm{exp}(-t/\tau_2),
    \label{eq:raman}
\end{equation}
where parameters for fused silica are $\tau_1 = \SI{12.2}{\femto\second}$ and $\tau_2=\SI{32}{\femto\second}$, yielding a Raman peak gain redshifted by $\SI{13.2}{\tera\hertz}$.

\begin{figure}[b!]
\centering\includegraphics[width=1\linewidth]{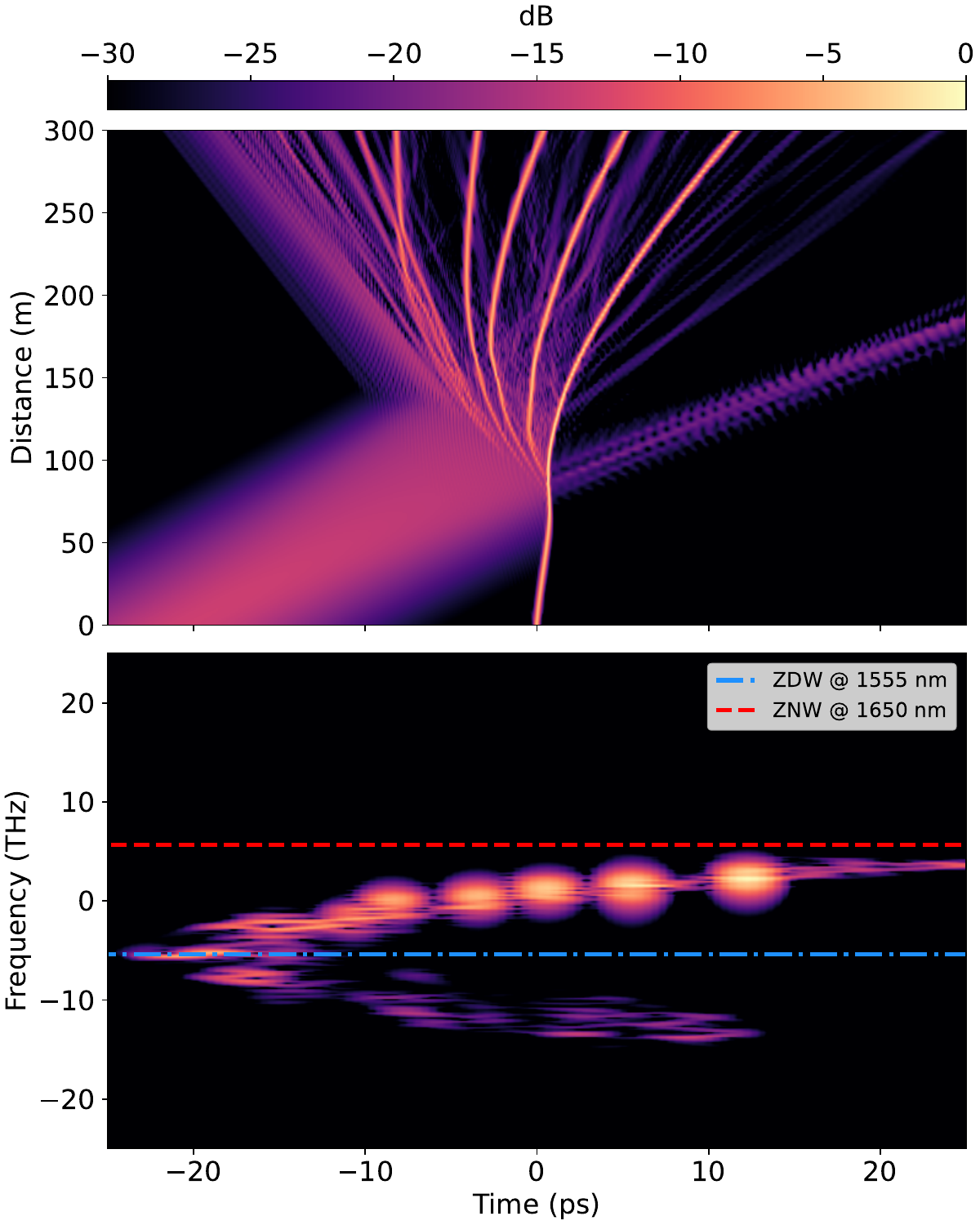}
\caption{Interaction between a dispersive pulse and a fundamental soliton (top) and spectrogram at the fiber output (bottom). The power is normalized to that of the launched soliton.}
\label{fig:evolspecgram}
\end{figure}

Figure~\ref{fig:evolspecgram} shows pulse evolution in time (top) and the spectrogram at the fiber output end (bottom). During the interaction numerous short pulses with peak powers ranging from \SI{70}{} to \SI{300}{\watt} are generated leading the initial soliton. These new pulses, that also experience Raman self-frequency shift, follow similar trajectories as the original soliton. The spectrogram reveals that the generated pulses lie between the ZNW and the ZDW where $\gamma(\Omega)\cdot\beta_2(\Omega) < 0$, i.e., a solitonic region. The soliton order $N^2 = \gamma(\Omega) P T^2/|\beta_2(\Omega)|$ was calculated for each of these pulses, whereby six fundamental solitons are readily singled out, with orders ranging from $N = 0.68$ to $N = 1.45$. \textit{Recall that soliton generation is not observed when neglecting the effect of Raman.} Furthermore, the order of the initial soliton was monitored and it remained nearly constant throughout propagation, indicating that it stays a fundamental soliton all along. This last observation confirms that new solitons are not generated by fission of the original soliton but rather through fission of the reflected wave.

\begin{figure}[t!]
\centering
\includegraphics[width=1\linewidth]{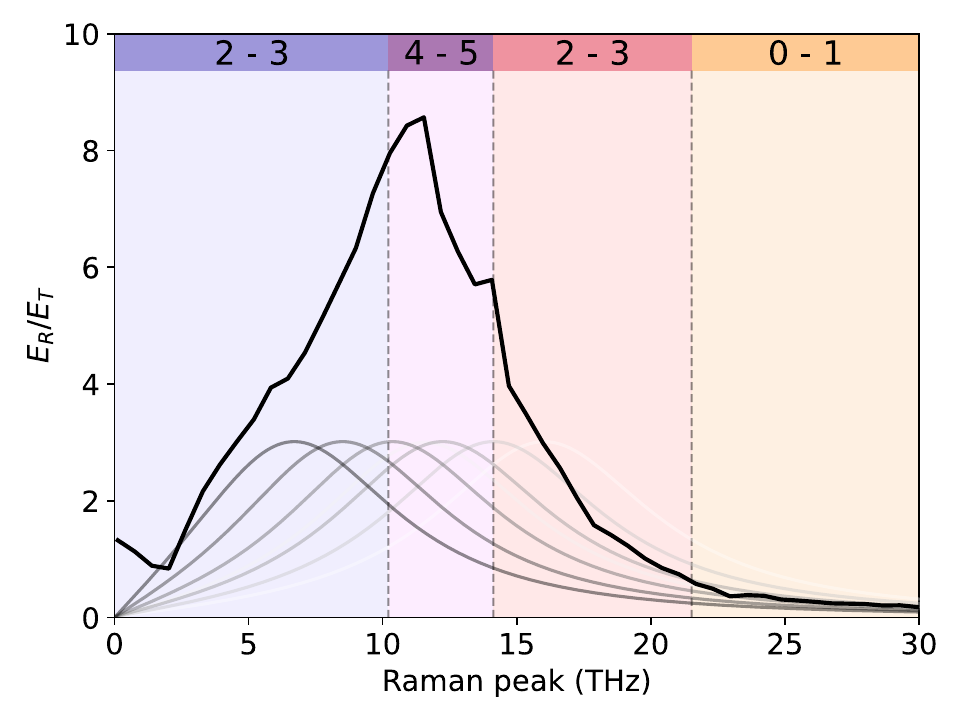}
\caption{Ratio of reflected to transmitted energy vs. the peak Raman gain frequency. Color-shaded areas display the number of generated solitons. Gray curves depict changes in the Raman gain spectrum as its peak is shifted.}
\label{fig:ramansweep}
\end{figure}

To further analyze the impact of Raman scattering on soliton generation, we performed 50 simulations using the same parameters as in Fig.~\ref{fig:evolspecgram}, while varying the frequency of the peak Raman gain by changing $\tau_1$ in \eqref{eq:raman}. Figure~\ref{fig:ramansweep} shows the ratio of reflected energy to transmitted energy vs. the frequency of the peak Raman gain, and the number of new solitons is shown in color-shaded areas. Note that the reflected energy is as high as 5x the transmitted energy when the peak Raman gain lies between \SI{7}{} and \SI{14}{\tera\hertz}. This can be explained as follows: When the dispersive pulse begins to interact with the soliton, a part of it reflects across a broad frequency range as a result of the soliton spectral shift. Whenever reflected spectral components fall within the peak Raman gain band provided by the remaining of the incident dispersive pulse, the reflected pulse is benefited from a resonant energy transfer, thus enhancing reflectivity and reducing transmitted energy, followed by its fission into multiple solitons. This is shown in Fig.~\ref{fig:ramansweep}, where more solitons are observed to be generated when the Raman peak lies close to \SI{13}{\tera\hertz}, thus coinciding with the portion of the spectral band of the reflected pulse that lies within the solitonic region.

\begin{figure}[t!]
\centering\includegraphics[width=1\linewidth]{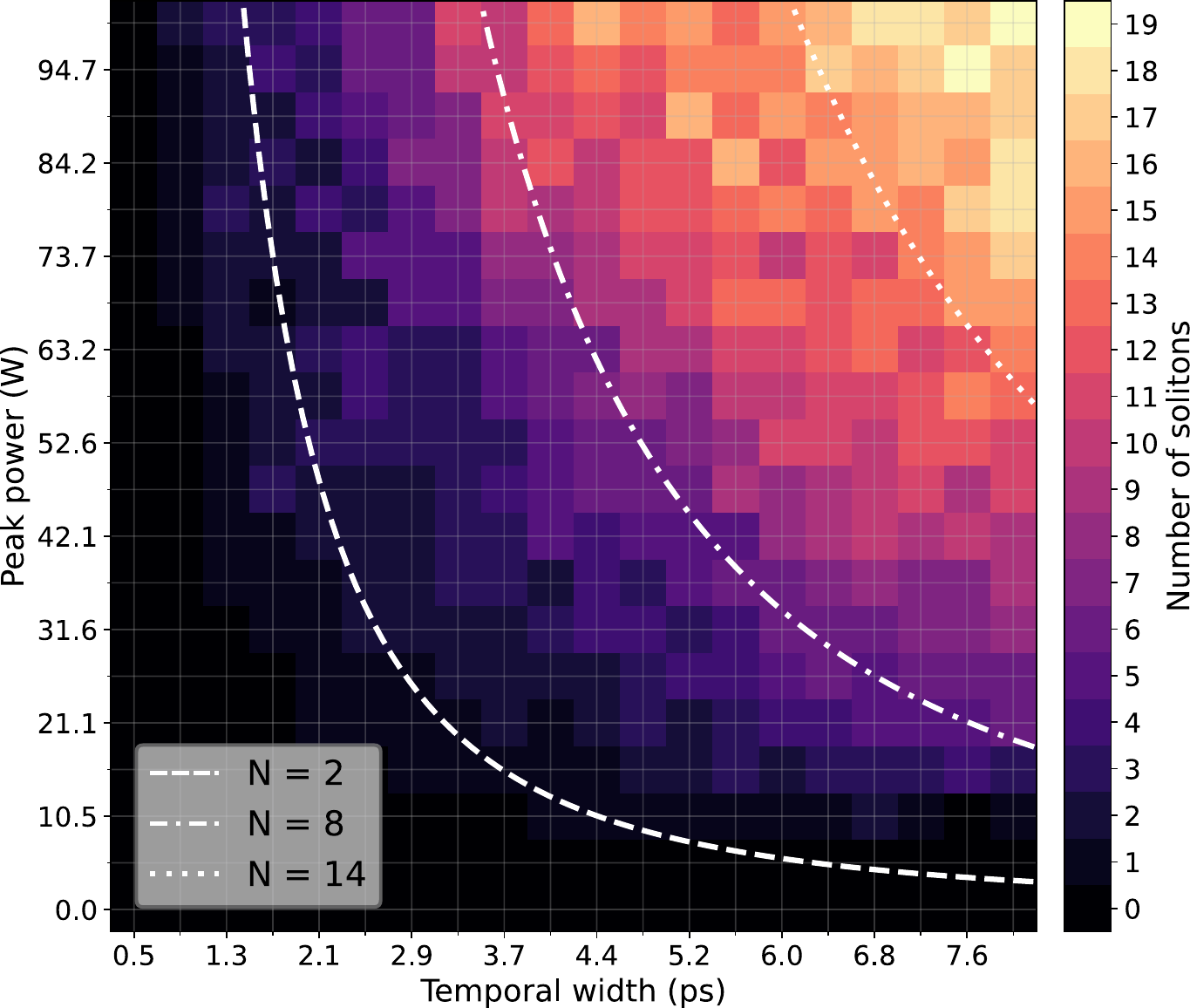}
\caption{Number of solitons vs. dispersive pulse power and time width. White curves correspond to $N = 2$ (dashed line), $N = 8$ (dashed-dotted line), and $N = 14$ (dotted line).}
\label{fig:sweep}
\end{figure}

Bearing in mind potential applications of the proposed scheme, we now turn our attention to the dependence of soliton generation upon controllable parameters of the dispersive pulse, such as power and time width. To this end we performed 400 simulations and, in each case, we determined the number of generated solitons by fitting each short intense pulse with a \textit{sech} function, whose central frequency was found by means of a Fourier transform of the peak under analysis. Finally, the order $N$ of each short pulse was estimated and it was deemed a soliton whenever $N > 0.5$. Figure~\ref{fig:sweep} shows the number of generated solitons as a function of both the time width and the peak power of the dispersive pulse. Close inspection reveals that while the number of new solitons does not vary smoothly, it does follow a distinct trend. We see that peak powers lower than \SI{6}{\watt} or time widths shorter than \SI{0.9}{\pico\second} produce no new solitons, as short weak pulses are unable to cause a significant spectral shift of the fundamental soliton, therefore seeding little reflected energy to the solitonic regime. Moreover, the short duration of the incident pulse restricts its overlap with the reflected pulse, thereby limiting the effect of Raman energy transfer. On the other hand, by increasing pulse duration or peak power more solitons are generated, as a result of a greater spectral shift of the soliton along with a significant overlap between incident and reflected pulses, producing up to 19 new solitons for a dispersive pulse of \SI{8}{\pico\second} and \SI{100}{\watt}.

While characteristics of the reflected pulse may be difficult to predict, a back-of-the-envelope calculation can provide a rough estimate. Indeed, if we assume that both its power and time width are proportional to those of the incident dispersive pulse, we can estimate its order as $N_R^2 = K \gamma(\Omega_\mathrm{R}) P_\mathrm{d} T_\mathrm{d}^2/|\beta_2(\Omega_\mathrm{R})|$ with $\Omega_\mathrm{R}$ the frequency of Raman peak gain with respect to the dispersive pulse, and $K$ a proportionality constant. White curves in Fig.~\ref{fig:sweep} show the relation between the peak power and time width for $N_R = 2$ (dashed), $N_R = 8$ (dash-dotted) and $N_R = 14$ (dotted), where the proportionality constant $K$ was determined for each case. As it can be observed, the number of new solitons is directly connected to the order of the reflected pulse, confirming that soliton generation stems from a fission process of the latter. Do note that the curve for $N_R = 2$ does not accurately represent conditions at low powers, as the spectral shift experienced by the original soliton turns out to be not sufficient to place the reflected pulse within the solitonic regime.

\begin{figure}[t!]
\centering
\includegraphics[width=1\linewidth]{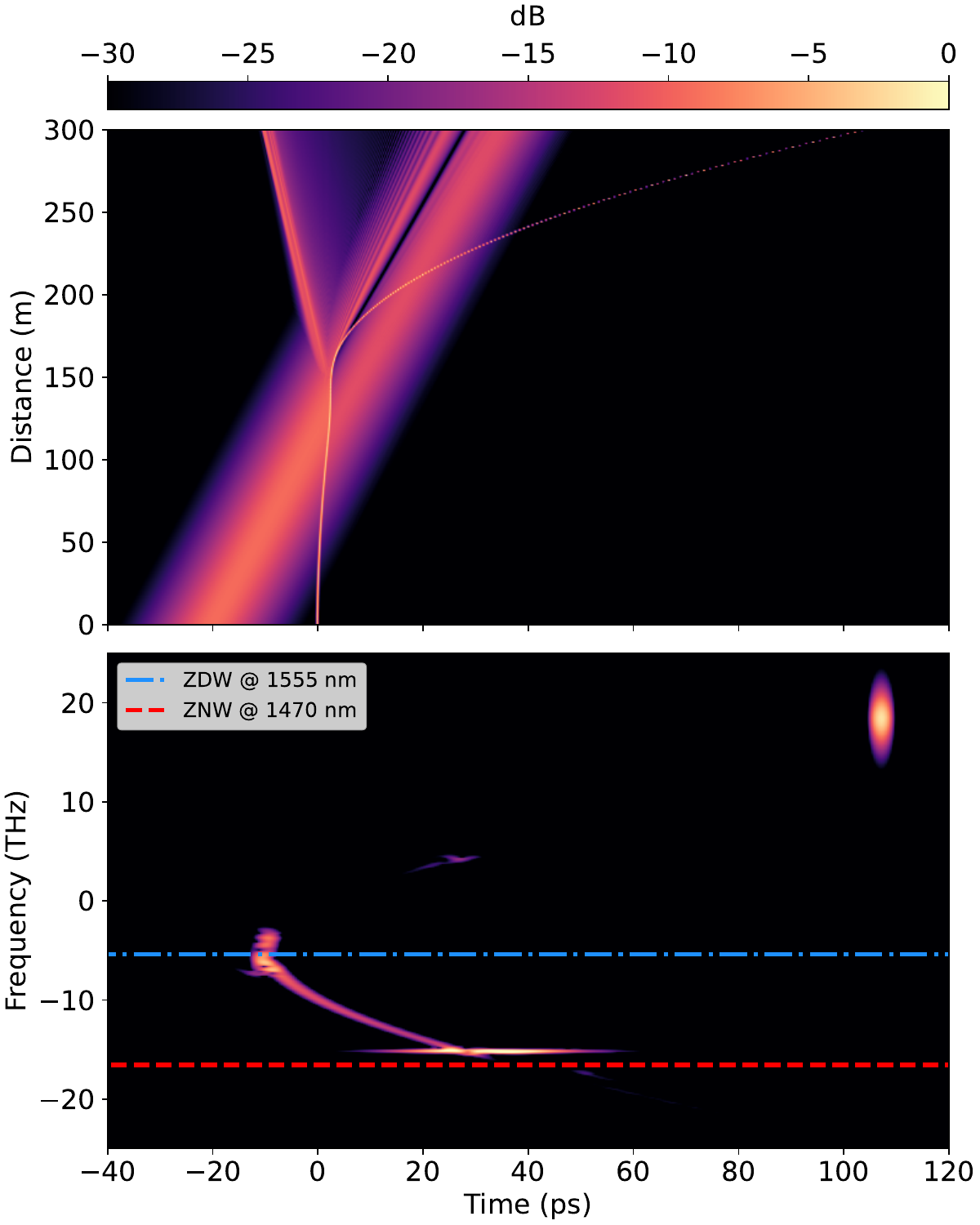}
\caption{Time evolution (top) and spectrogram (bottom) for ZNW @ \SI{1470}{\nano\meter}. No soliton generation is observed}
\label{fig:ZNW}
\end{figure}

The position of the ZNW is critical for the generation of new solitons, as shown in Fig.~\ref{fig:ZNW}, where no such process is observed when ZNW $=\SI{1470}{\nano\meter}$. The influence of the zero-nonlinearity wavelength is manifested in two different ways. First, the zero nonlinearity has been shown to act as a barrier for the Raman-induced redshift of the fundamental soliton~\cite{Bose:16_Implications_ZNW}. A ZNW at \SI{1650}{\nano\meter} minimizes the soliton's redshift before its interaction with the dispersive pulse, maintaining the group-velocity difference between pulses small and, hence, preserving a high reflectivity throughout~\cite{Plansinis:15_Temporal_Analog}. Second, the minimal frequency shift of the soliton prior to the interaction ensures that the blueshift it suffers during the collision is sufficient for the reflected wavelengths to cross the ZDW into the solitonic region, as described by \eqref{fig:recoil}.

In conclusion, we studied the generation of new solitons resulting from the temporal reflection of a weak dispersive pulse on a soliton. We showed examples of such a schema, where this process depends on the critical contributions from both the soliton spectral shift, by enabling the reflected pulse to fall within a solitonic spectral region, and the resonant energy transfer from the incident to the reflected pulse through Raman scattering. Such combination makes for the generation of numerous new solitons even with a modest power of the dispersive pulse. 
We also addressed the role played by the zero-nonlinearity wavelength, which, depending upon its relative position to that of the launched soliton, may limit its Raman induced self-frequency shift, preserving a high reflectivity and maximizing the effect of the blueshift during the collision.
Finally, our analysis of dispersive-pulse parameters reveals that the number of generated solitons can be easily controlled. We believe results put forth in this work may contribute general guidelines for the generation of short and intense pulses for various applications in the realm of photonics.

\begin{backmatter}
\bmsection{Funding} Consejo Nacional de Investigaciones Científicas y Técnicas
(PIP 2021 GI - 11220200100566CO).

\bmsection{Disclosures} The authors declare no conflicts of interest.

\bmsection{Data availability} No data were generated or analyzed in the presented research.

\end{backmatter}

\bibliography{bibliography}

\bibliographyfullrefs{bibliography}

\end{document}